Full article

# Monitoring dynamic collagen reorganization during skin stretching with fast polarization-resolved SHG imaging.

*Guillaume Ducourthial[1], Jean-Sébastien Affagard[2], Margaux Schmeltz[1], Xavier Solinas[1], Maeva Lopez-Poncelas[2], Christelle Bonod-Bidaud[3], Ruth Rubio-Amador[3], Florence Ruggiero[3], Jean-Marc Allain[2,4], Emmanuel Beaurepaire[1], Marie-Claire Schanne-Klein[1]\**

\*Corresponding Author: E-mail: marie-claire.schanne-klein@polytechnique.edu

[1] LOB, Ecole Polytechnique, CNRS, INSERM, 91128 Palaiseau, France

[2] LMS, Ecole Polytechnique, CNRS, 91128 Palaiseau, France

[3] Institut de Génomique Fonctionnelle de Lyon, ENS-Lyon, CNRS, Université de Lyon, 69007 Lyon, France

[4] Inria, Université Paris-Saclay, 91128 Palaiseau, France

Keywords: multiphoton microscopy, polarization, biomechanics, collagen, skin

## Abstract

The mechanical properties of biological tissues are strongly correlated to the specific distribution of their collagen fibers. Monitoring the dynamic reorganization of the collagen network during mechanical stretching is however a technical challenge because it requires mapping orientation of collagen fibers in a thick and deforming sample. In this work, a fast polarization-resolved SHG microscope is implemented to map collagen orientation during mechanical assays. This system is based on line-to-line switching of polarization using an electro-optical modulator and works in epidetection geometry. After proper calibration, it successfully highlights the collagen dynamic alignment along the traction direction in *ex vivo* murine skin dermis. This microstructure reorganization is quantified by the entropy of the collagen orientation distribution as a function of the stretch ratio. It exhibits a linear behavior, whose slope is measured with a good accuracy. This approach can be generalized to probe a variety of dynamic processes in thick tissues.

**1. Introduction**

Collagen is the main component of the extracellular matrix and is organized as a network of fibrils in many tissues.[1, 2] The distribution of this fibrillar collagen at the microscopic scale determines to a large extent the biophysical and biomechanical properties of tissues. However, while the mechanical behavior of tissues such as skin have been thoroughly characterized at macroscopic scale, the precise relationship between these macroscopic properties and the microstructure of collagen fibers is still a debate because of technical issues in acquiring multiscale experimental data.[3-8] *In situ* characterization of the collagen distribution in intact tissues is therefore a major issue to decipher the biomechanics of healthy and pathological tissues.

To address this issue, we have recently implemented a multiscale setup combining a traction device and Second Harmonic Generation (SHG) imaging.[9, 10] SHG microscopy has indeed become the reference technique for 3D imaging of collagen at the microscopic scale.[11] SHG highlights collagen fibers with high specificity and sensitivity in intact tissues without requiring any staining.[12] Imaging of the collagen organization can be further improved by using polarization-resolved SHG microscopy (P-SHG).[13] This technique combines the sensitivity of polarimetry to molecular orientation and order with the specificity of SHG for fibrillar collagen. P-SHG measurements therefore provide additional information about the orientation and hierarchical structure of collagen fibers in the tissue.[14-19] However, P-SHG microscopy requires the acquisition of an image sequence recorded with a series of linear incident polarizations and is therefore a slow technique. Reliable P-SHG-based determination of the collagen reorganization during dynamic processes is therefore impossible if the sample exhibits distortions on the time scale needed to perform measurements with different polarizations. A number of strategies have been developed to overcome this limitation and have proven efficient in specific samples.[20-25] Nevertheless, most of these approaches are based on polarization analysis of forward SHG signals, which is not applicable to thick scattering tissues, or they use only two incident polarizations for SHG imaging, which limits their accuracy in tissues with disordered collagen. A robust method enabling fast epidetection of P-SHG signals is therefore needed to monitor collagen dynamic reorganization in thick deforming tissues.

In this paper, we describe an original fast P-SHG imaging setup based on line-to-line switching of polarization using an electro-optical modulator (EOM). As the EOM is located outside the SHG microscope, a careful calibration is performed to obtain reliable P-SHG measurements. We use this P-SHG system to map the dynamic reorganization of the collagen network in *ex vivo* murine skin dermis during mechanical assays. We show that our fast setup enables monitoring collagen orientation even in this deforming sample and provides an accurate measurement of collagen reorganization during skin stretching.

## 2. Materials and methods

### 2.1. Fast P-SHG based on line-to-line EOM polarization switching

Multiphoton imaging is performed using a custom-built upright laser scanning microscope.[9] Excitation is provided by a femtosecond laser set at 860 nm (80MHz, 100fs, Mai-Tai, Spectra-Physics), with typically 20-30 mW excitation power at the focus of a water-immersion 20×, 0.95 NA objective lens (Olympus), resulting in 0.6 μm (lateral) × 3.2 μm (axial) resolution measured near the sample surface. Multiphoton signals are epi-detected using photon-counting photomultiplier tubes (P25PC, Electron Tubes) and appropriate dichroic mirrors and spectral filters.

An Electro-Optic Modulator (EOM) (M350-210-02, Conoptics) is used to achieve fast rotation of the incident linear polarization as in the pioneering work by Stoller et al.[13] The incident polarization is set at 45° from the EOM axis using a half-waveplate (ACWP-700-1000-06-2, CVI) and the EOM is followed by a quarter-waveplate (ACWP-700-1000-06-4, CVI) oriented at 45° from the EOM axis (**Figure 1.A**). The resulting polarization orientation angle is then directly proportional to the input EOM voltage and can be tuned in the range 0°-170° at 860 nm. Polarization switching is achieved in less than 11 μs using a fast power supply (302A, Conoptics). It is synchronized with the microscope line scanning as displayed in Figure 1.B.

A quarter-waveplate is inserted at the back pupil of the objective lens to compensate for the ellipticity introduced by the microscope optical components. After optimization of all waveplate orientations, the ellipticity $E_{min}/E_{max}$ is less than 10 % for all polarization angles and the power variation with the polarization angle is less than 1% (see Figure S1 in Supporting Information). Nevertheless, for any given setting, the scanning process causes a slight variation of the polarization angle across the field of view. This effect is calibrated by imaging a homogeneous fluorescent plate with a rotating polarizer inserted at the back pupil of the objective lens and the P-SHG images are corrected accordingly (see Supplementary method and Figure S2 in Supporting Information).

Alternatively, slow P-SHG imaging can be performed by use of a motorized half-waveplate at the back pupil of the objective as reported previously.[26] In this case, as mechanical rotation is slow, polarization orientation changes are performed between two successive images instead of between lines.

## 2.2. Traction assays in murine skin under P-SHG microscope

All animal experiments are performed under animal care procedures in accordance with the guidelines set by the European Community Council Directives (86/609/EEC). All experimental procedures are approved by the Direction of the Veterinary Service of Rhone Department (DDSV, Lyon, France).

Reorganization of the collagen network is monitored in *ex vivo* murine skin dermis during traction assays as previously described.[9, 27] Eight wild-type mice (129sv) are sacrificed by cervical dislocation at one month. Skin from the back is depilated, the epidermis is removed using 3.8 % ammonium thiocyanate [28] and the remaining dermis is cut into a dog-bone shape along the antero-posterior direction to ensure homogeneous uniaxial tensile load in the central tested portion.[9] Skin samples are attached to a custom-built uniaxial traction device using metallic jaws and inserted in place of the microscope stage, with the papillary dermis up facing the objective lens. Symmetric stretching is applied at a slow strain rate (typically traction speed of 2 $\mu m.s^{-1}$) to monitor the same region of interest (ROI) during all the experiment while measuring the force every second. Stretch is maintained constant during multiphoton imaging, resulting in an incremental loading path with steps of 0.05 stretch ratio. No preconditioning is applied before stretching. Immersion gel (Lacrygel, Europhta) is used to ensure optical contact with the objective lens and to prevent skin dehydration during experiments.

$375 \times 375 \times 50$ $\mu m^3$ image stacks are epi-collected at 100 kHz pixel rate, using 0.5 $\mu m$ pixel size and 2 $\mu m$ axial steps. 18 different incident polarizations (0-170° every 10°) are used for P-

SHG. The total acquisition time of all polarizations for a line of 750 pixels using EOM switching is 200 ms.

Mechanical data are processed as previously described.[9, 27] Briefly, the global stretch ratio λ is obtained as the imposed length between the jaws divided by the initial length, and the nominal stress as the measured force divided by the initial skin section. The tangent modulus is then obtained as the slope of the linear part of the nominal stress/stretch curve (see Supporting Information).

**2.3. P-SHG theory and image processing**

The SHG response of collagen fibers is described by a second order nonlinear susceptibility tensor $\underline{\underline{\chi}}^{(2)}$. Considering that the collagen fibers exhibit a cylindrical symmetry and that the Kleinman symmetry applies,[13] there are only 2 independent components: $\chi_{xxx}$ and $\chi_{xyy} = \chi_{xzz} = \chi_{yxy} = \chi_{yyx} = \chi_{zxz} = \chi_{zzx}$ where x is the fiber axis. Derivation of the P-SHG signal intensity from a collagen fiber oriented at an angle φ in the image plane XY then requires to change from the fiber frame xyz to the microscope reference frame XYZ and to calculate the SHG response to a linearly polarized excitation at an angle θ to the X axis (see inset in Figure 1(A)). It writes in the plane wave approximation, without any polarization analysis of the SHG signal because of the epidetection configuration:

$$I_{SHG}(\theta) = K\left[|\chi_{XXX} \cos^2(\theta - \varphi) + \chi_{XYY} \sin^2(\theta - \varphi)|^2 + |\chi_{XYY} \sin(2(\theta - \varphi))|^2\right] \quad (1)$$

Equation (1) also applies to fibers lying out of the image plane and projected in the image plane at an angle φ to the X axis, since the out of plane dependence is included in the susceptibility components in the XYZ frame. The parameter K includes various geometrical parameters and the square of the excitation laser intensity. Note that this analytical approach does not take into account the axial components of the excitation field due to strong focusing.[29] Indeed, we have recently demonstrated that strong focusing does not affect the determination of the in-plane orientation φ of collagen fibers, while it strongly affect other parameters, notably the ratio of the two susceptibility components.[26]

It is convenient to write Eq. (1) as a function of its Fourier components:[13, 30]

$$I_{SHG}(\theta) = K[A + B\cos(2(\theta - \varphi)) + C\cos(4(\theta - \varphi))] \qquad (2)$$

Where A, B and C are related to the susceptibility components of the collagen fiber. In this study, we are only interested in the in-plane orientation φ of the collagen fiber. We use a FFT algorithm to determine the A, B and C complex parameters in every pixel and the orientation φ is obtained from the phases of B and C.[26] The reliability of this determination is assessed by a coefficient of determination, so-called $R^2$ parameter, that compares the experimental data and the curve obtained from the A, B and C parameters extracted with FFT. Angles determined with $R^2$<0.5 are skipped in the quantitative analyses, which ensures an angular accuracy of 5°. This accuracy is estimated by the difference between the first angle φ extracted from B and the second angle φ extracted from C. The orientation maps are displayed using a HSV Look-Up-Table: the Hue (H) indicates the collagen orientation φ and the brightness (V) the $R^2$ parameter: V=1 for $R^2 \geq 0.5$ and V $\in$[0;1] for $R^2 \in$[0;0.5].

**2.4 Analysis of the reorganization of the collagen fiber network**

The resulting map of fiber orientations is further processed to plot a histogram of the orientations within the field of view. Our previous studies have shown that tissue relaxation is negligible during the acquisition of the P-SHG image,[27] so that we can safely quantify parameters from this histogram, namely calculate the entropy S of this orientation distribution. S is defined as the usual statistical entropy, using angular bins of 1° or 10°, and is normalized to be independent of the number of bins nb(θ):

$$S = \frac{-1}{\ln[nb(\theta)]} \sum_{\theta=-90°}^{90°} p(\theta)\ln[p(\theta)] \qquad (3)$$

Where p(θ) is the normalized number of pixels with the orientation θ in the image, which is directly obtained from the orientation histogram. The entropy S then assesses the degree of orientation disorder in the field of view: S=1 for a fully isotropic orientation distribution and decreases towards 0 for thinner orientation distribution. This analysis is performed for the six consecutive 2D images exhibiting the maximal SHG intensity in the P-SHG z-stack. The mean ± SD (standard deviation) entropy is calculated and plotted as a function of the stretch ratio for every skin sample. Linear fitting gives the slope of the entropy variation that characterizes the reorganization of the collagen fiber network upon stretching.

Alternatively, the entropy is computed from orientation maps obtained by image processing of the SHG images using the morphological filtering method reported previously.[9] Briefly, all the SHG images obtained for different linear excitations are summed up to mitigate the impact of polarization versus fiber orientation on the SHG intensity. The local orientation of the fibers is then extracted in every pixel using morphological filtering by a rotating linear structuring element (19 pixels/9.5 μm long). This method provides histograms with 10° bins.[9]

## 3. Results

**3.1 Fast P-SHG imaging allows orientation measurements in dynamic samples**

P-SHG imaging was performed in the dermis of eight wild-type mice during traction assays and the collagen orientation was processed pixel-wise using Eq. (2). **Figure 2** displays the orientation map for the WT4 mouse at 1.05 stretch ratio, as well as the SHG image obtained as the summation of all images acquired with the set of different linear incident polarizations (sum-SHG image, Figure 2.A). This sum-SHG image allows the visualization of the collagen fibers regardless of their orientation in the image plane, in a similar way as the SHG image obtained by use of circularly-polarized excitation (circular-SHG image, Figure 2.C). Nevertheless, the circular-SHG image is recorded much faster (8.4 s for 750x750 pixels at 100 kHz pixel rate) than the sum-SHG image that requires the acquisition of 18 polarizations for every line (18 x 8.4 s = 150 s). Accordingly, motion artefacts are observed in the sum-SHG image, not in the circular-SHG one. Nevertheless, these line-to-line motion artefacts do not impede the determination of the collagen orientation because P-SHG images are acquired using fast rotation of polarization synchronized with line scanning. The same line is recorded sequentially with 18 different incident linear polarizations using fast polarization switching by the EOM device, then the scanner moves to the next line. This method enables fast recording of P-SHG lines (200 ms), even if the full image is recorded much slower (200ms x 750 lines = 150s, see above). Accordingly, Figure 2.B shows that the collagen orientation is satisfactorily processed from the P-SHG images even in the regions with motion artefacts.

Note that faster motion artefacts in the observed tissue would not be removed by this method and would necessitate decreasing the number of pixels in order to decrease the acquisition time. Nevertheless, such fast and strong movements are only observed sparsely at the very beginning of the traction assay, probably due to unfolding of skin wrinkles, which means that they do not affect the linear part of the collagen reorganization (see next section).

**3.2. Fast P-SHG imaging of collagen fiber network reorganization**

Fast P-SHG was used to assess the dynamic reorganization of collagen fibers in *ex vivo* murine dermis using sequential 0.05 stretching steps. Force measurements were processed to plot the stress-stretch curve and derive the tangent modulus as shown in Supporting Information (Figure S3). We obtained a J-shaped curve as in our previous studies,[9] which means that the increased duration of the traction assays due to polarization resolution did not strongly affect the mechanical response of the dermis at macroscopic scale. The tangent modulus is slightly larger in this study: 2.1 ± 0.2 MPa (N=8) than in the previous one: 1.6 ± 0.1 MPa (N=31), but the difference cannot be considered as statistically significant (p=0.07 using Wilcoxon test).[10]

**Figure 3.A** displays typical sum-SHG images at a few stretch ratios, the loading axis being the horizontal direction. All these images display a similar pattern of elliptical features without any SHG signal, which correspond to the positions of the hair follicles. It shows that we successfully imaged the same region of interest at all stretch ratios. The deformation of this pattern upon stretching enables the computation of a local stretch ratio, in excellent agreement with the imposed global one (data not shown).[9] It shows that the dermis response is homogeneous and validates the mechanical assays.

Figure 3.A also shows that the collagen fibers gradually align along the traction direction at increasing stretch ratios. This effect is better revealed in Figure 3.B, which displays the orientation maps obtained from the P-SHG images. The maps at low stretch ratios display various colors corresponding to various orientations, while the maps at high stretch ratios gradually turn red that is the color of the horizontal traction direction.

**Figure 4.A** displays the orientation histograms obtained from these orientation maps at increasing stretch ratios. The collagen fibers show a main orientation close to the traction axis, which corresponds to the antero-posterior direction, even at low stretch ratio, as already reported.[27] The observed peak is quite wide at low stretch ratios and becomes gradually thinner at higher stretch ratios. Accordingly, the entropy of these orientation distributions decreases when the stretch ratio increases, as shown in Figure 4.B. Note that the entropy is sufficient to fully characterize the distribution here as the mean orientation of the collagen fibers is aligned with the stretching direction. We observe a linear decrease, which is preceded by a plateau at low stretch ratios in some skin samples. This observation is in agreement with our previous experiments that used morphological filtering to extract orientation maps from SHG images acquired with circular polarization.[9] It indicates that the increased duration of the experiment due to polarization resolution does not strongly affect the reorganization of the collagen microstructure in murine skin samples. The slope obtained by linear fitting then

provides a quantitative measure of this reorganization, as summarized in **Table 1** for the eight skin samples.

### 3.3. Comparison with morphological filtering

For the sake of comparison, we also processed the sum-SHG images using morphological filtering as previously reported.[9] The corresponding orientation maps, orientation histograms and entropy evolution as a function of the stretch ratio are depicted in the Figure S4 (Supporting Information), and the resulting slopes for the eight murine samples are summarized in Table 1. These orientation results show the same trends as the ones obtained from P-SHG, but the orientation maps derived from the sum-SHG images show more different orientations and the histograms are wider and exhibit higher background. As a consequence, the values of the entropy and of the slope of its evolution upon stretching are different from the ones obtained from P-SHG. [31, 32]

We also compared the respective precision of P-SHG and morphological filtering by looking at the accuracy of the linear fitting of the entropy decrease upon stretching. This accuracy is assessed by calculating the R² parameter of the linear fit. It is slightly larger for P-SHG (0.95 ± 0.02) than for morphological filtering (0.88 ± 0.03).

### 4. Discussion and conclusion

In this study, we implement a fast P-SHG microscope based on an EOM, following the pioneering work by Stoller et al.[13] We take advantage of the fast switching rate of the EOM to synchronize the rotation of incident linear polarization with line scanning and record XPY images. Our microscope advantageously enables imaging of thick scattering samples because it is based on epidetection without polarization analysis. It may easily be generalized to circular polarizations as well. Note however that a careful calibration of the polarization states is necessary because the EOM device is conveniently installed outside of the microscope, resulting in the propagation of the incident polarization states across potentially distorting components of the microscope. Nevertheless, once the calibration is completed, this fast P-SHG microscope is a robust and application-friendly setup.

Importantly, the acquisition duration depends on several parameters here: the number of pixels per line, the pixel dwell time and the number of incident polarization states. While the total acquisition time of a XPY image is longer than with circular-polarization (150 s versus 8.4 s for 750 x 750 pixels), the relevant time here is the one for line imaging: 200 ms for 750 pixels. This acquisition duration can be easily minimized by decreasing the number of pixels in every line, either using larger pixels (faster scanning) or imaging smaller areas. For instance, lines of 100 pixels instead of 750 pixels would result in a P-SHG acquisition time of 27 ms per line instead of 200 ms. This acquisition time is only moderately longer than the one relevant for morphological filtering in one pixel, that is the acquisition time of the window used for filtering (in our case, 19 x 19 pixels): 5.4 ms. Some studies have proposed to further reduce the acquisition time by using only 2 or 3 polarization states either for the excitation or for the analysis of circularly-excited SHG.[21, 22, 24] While this approach may be efficient in samples with simple orientation distribution, it cannot provide accurate orientation maps in samples with more complex orientation distribution.

In practice, the pixel dwell time and the number of incident polarization states cannot be easily decreased because they determine the signal to noise ratio in the P-SHG images. Indeed, the accuracy of the determination of the collagen orientation and of other parameters extracted from P-SHG images is related to the total number N of detected photons.[33, 34] This number N scales as the pixel dwell time multiplied by the number of incident polarization states. It varies in every biological tissue: it mainly depends on the maximum intensity usable for imaging without photo-damaging the tissue and on the collagen density and alignment in the tissue, which determines the efficiency of the SHG process. It is also smaller in depth because scattering and aberrations may decrease the excitation intensity. Practically, in our murine skin dermis samples, we safely use 20-to-30 mW excitation power and detect approximately N=500 photons in total per pixel. This corresponds to an accuracy of 3° on the orientation determination, which is slightly degraded to 5° (see R² threshold in section 2.2) by the residual ellipticity of the excitation and mostly by possible polarization distortions due to heterogeneities within the skin sample.[26] Compromises on this accuracy would enable to speed up the acquisition, for instance 7 ms for one line of 100 pixels using 9 polarizations instead of 18 ones and half pixel dwell time (5 µs). Nevertheless, given the order of magnitude of SHG signals detected in collagen tissues, the minimal pixel dwell time in a point-scanning microscope is a few µs. It means that in collagenous tissues, the main limitation to decrease the acquisition duration is the relatively low SHG signal, not the polarization switching rate. Smart approaches to achieve video-rate polarization switching have been reported recently and proved efficient in samples exhibiting higher signals.[23, 25, 35] Nevertheless, in complex biological samples such as skin dermis, such imaging rates require to rely on subsequent frame averaging to achieve a sufficient signal to noise ratio and result in a similar acquisition duration as the one we obtained for full images. In that respect, our strategy based on fast line-to-line P-SHG imaging is a good compromise: it is slow enough to detect small endogenous SHG signals and fast enough to determine collagen orientation in dynamic samples, as demonstrated in Figure 2. Moreover, it works in thick samples where epi-detection is mandatory, which is not the case for many other approaches. [22-25, 35] It may be further improved by synchronizing the EOM with pixel imaging in order to image tissues with faster motion. In case of even faster tissue deformations, processing SHG images acquired with circular polarization, short pixel dwell time and low sampling may be the only possible technique to determine collagen orientation, probably at the expense of the accuracy.

We use here this fast P-SHG microscope to characterize the reorganization of the collagen network in *ex vivo* murine dermis upon stretching. We previously performed similar experiments using morphological filtering to determine the collagen orientation from SHG images, and our results in wild-type and genetically-modified mice challenged the usual interpretation of the microstructural origin of the skin macroscopic biomechanical behavior.[27] Indeed, as we have previously reported [9, 10, 27], the collagen fibers do not align with the direction of traction in the non-linear part, but they do so in the later linear part of the stress/stretch curve. These observations are in contradiction with the classical assumption of the role of collagen in skin mechanical properties.[3-5] Nevertheless, advanced modeling of the collagen network response requires measurements of the collagen reorganization with a good spatial accuracy that is pixel-wise measurements, while methods based on image processing probe a larger scale. Indeed, our method based on morphological filtering of sum-SHG images provides a pixel-wise orientation map, but it probes in practice the collagen orientation at the scale of the rotating linear structuring element (here 19 pixels, that is approximately 10 µm). It is the case for all image processing techniques aiming at determining orientations, which always require to process a window around the pixel under study.[31, 32] In contrast, P-SHG effectively probes the collagen orientation at the pixel size (here 0.5 µm). P-SHG measurements are therefore required to improve the spatial accuracy of collagen orientation measurements during traction assays in the framework of our multiscale study.

Accordingly, we successfully image the reorganization of the collagen network in murine dermis using fast P-SHG imaging. We obtain thin orientation histograms with low background that show qualitative similarities to the wide histograms with high background obtained by morphological filtering. The observed quantitative differences are attributed to the different scales probed by the two methods. Moreover, both methods have to compromise between (i) the accuracy of the orientation maps (number or angular bins, pixel size and binning, $R^2$ filtering…) and (ii) the number of pixels with determined orientation and consequently the statistical significance of these orientation maps. This compromise does not imply the same parameters in both methods and is therefore difficult to adjust in the same manner, while the set of parameters used in every method may change the quantitation of the results. For instance, the results obtained from P-SHG images using 1° and 10 ° angular bins for the calculation of the entropy are quantitatively different. Nevertheless, the results obtained by P-SHG (using 1° or 10° bins) and by morphological filtering reveal the same trends, as shown by the last columns of Table 1: the ratio of the entropy slopes obtained by both methods exhibits low variation for the eight skin samples. Notably, these entropy slopes obtained by P-SHG are determined with a better accuracy than the ones obtained by morphological filtering. Note however that this conclusion holds for dermis, a tissue with a dense network of collagen fibers, but the situation may be different in other types of tissues, for instance in the case of sparsely distributed straight collagen fibers, which orientation is accurately determined by SHG image processing.[32, 36] These results are therefore a proof of feasibility of fast P-SHG imaging of skin microstructure reorganization during mechanical assays. Systematic analysis of more skin samples should therefore enable the determination of accurate orientation histograms and enable detailed comparison with multiscale mechanical models.

In conclusion, we report here on a new fast P-SHG microscope based on EOM polarization switching synchronized with line imaging. This method is fast enough to obtain reliable orientations in dynamic collagen samples, while slow enough to get a good SHG signal to noise ratio in thick collagenous tissues. We demonstrate that it enables accurate quantitation of dynamic collagen reorganization in murine skin dermis during stretching biomechanical assays. This approach can in principle be generalized to other multiphoton modes of contrast and enable fast imaging of a variety of orientation-sensitive processes.

**Supporting Information**

Additional supporting information may be found in the online version of this article at the publisher's website.

**Supplementary method: fluorescence-based calibration.**

**Figure S1.** Variation of incident (a) ellipticity and (b) power as a function of polarization angle. The ellipticity is defined as the ratio of electric fields minimum to maximum. The incident power is normalized to the maximum power.

**Figure S2.** Orientation shift of incident polarization in the microscope field of view (500 × 500 µm²). The calibration is performed in every pixel (320 × 320).

**Figure S3**. **Mechanical response of skin sample WT15** (same as in Figures 3 and 4). The nominal stress (measured force divided by the cross-section) is plotted as a function of the stretch ratio (imposed length divided by the initial length of the sample). The tangent modulus is obtained as the slope of the linear part.

**Figure S4. Reorganization of the collagen fibers network upon stretching** in skin sample WT15 (same as in Figures 3 and 4). A: Orientation maps at 1.1, 1.2, 1.3 and 1.4 stretch ratios processed from the sum-SHG images using morphological filtering. Look-up table: HSV, where H codes for the orientation and V for the intensity below $I_{max}/2$ (V=1 for $I>I_{max}/2$, V$\in$[0;1] for I$\in$[0;$I_{max}/2$]). Scale bar: 50 µm. B. Orientation histograms obtained from morphological filtering using 10° bins at increasing stretch ratios. The baseline is shifted upwards for increasing stretch ratios. C: Variation of the normalized entropy of the orientation distributions in B as a function of the stretch ratio. The black squares corresponds to the experimental data: mean ± SD calculated on the six 2D SHG images exhibiting the maximal signal; the solid line corresponds to linear fitting.

**Acknowledgements**

The authors acknowledge financial support from Agence Nationale de la Recherche under contracts ANR-10-INBS-04 (France BioImaging), ANR-11-EQPX-0029 (Morphoscope2) and ANR-13-BS09-0004-02 (Metis).

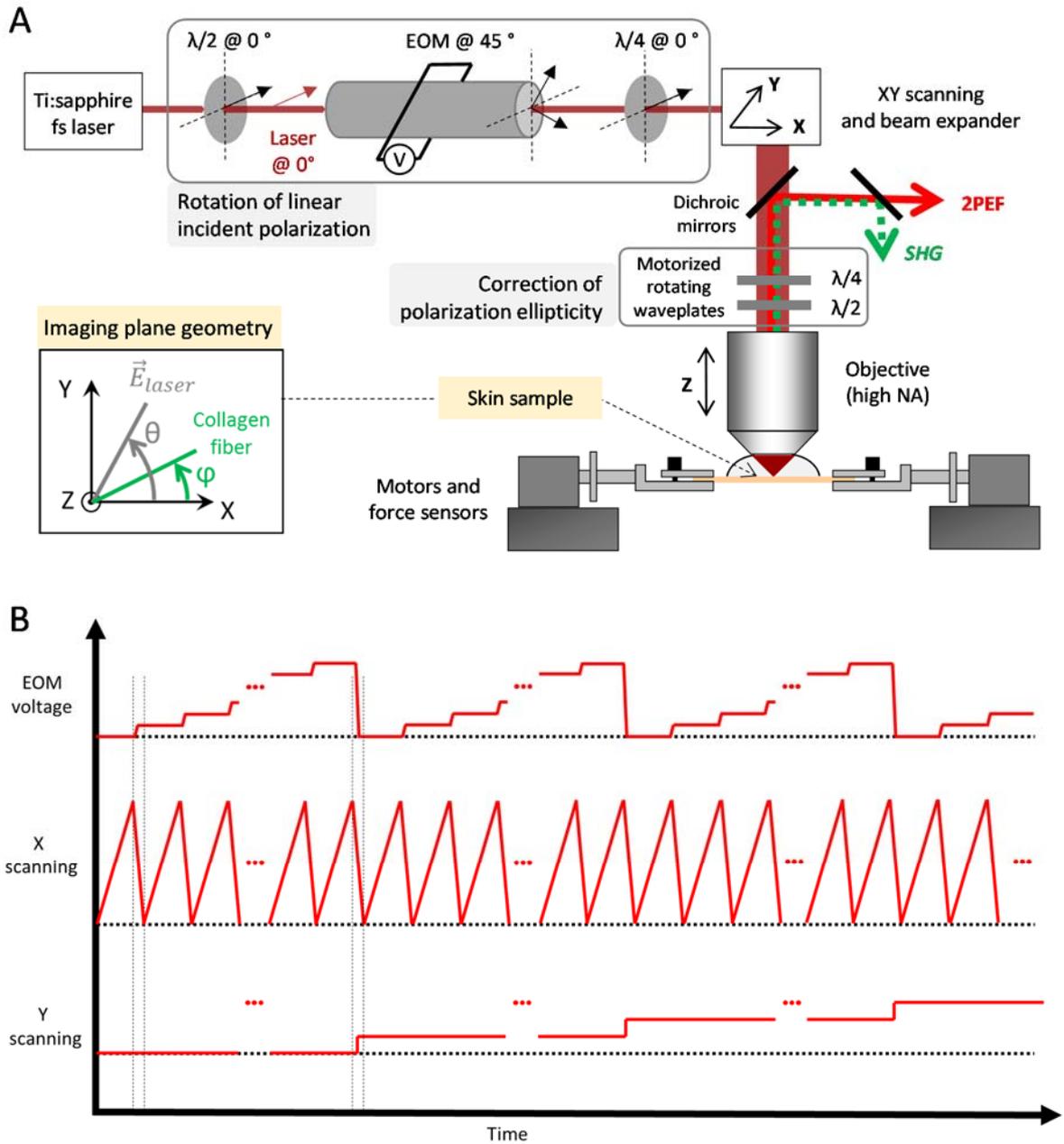

**Figure 1**. Experimental setup. (A) Fast P-SHG microscope using an EOM and a quarter-waveplate to rotate the incident linear polarization. The imaging plane geometry is shown in the insert. (B) Timing diagram. EOM voltage is synchronized with line scanning (x direction). The next line is scanned (y scanning) when the first line has been scanned for all EOM voltages (*i.e.* all polarizations).

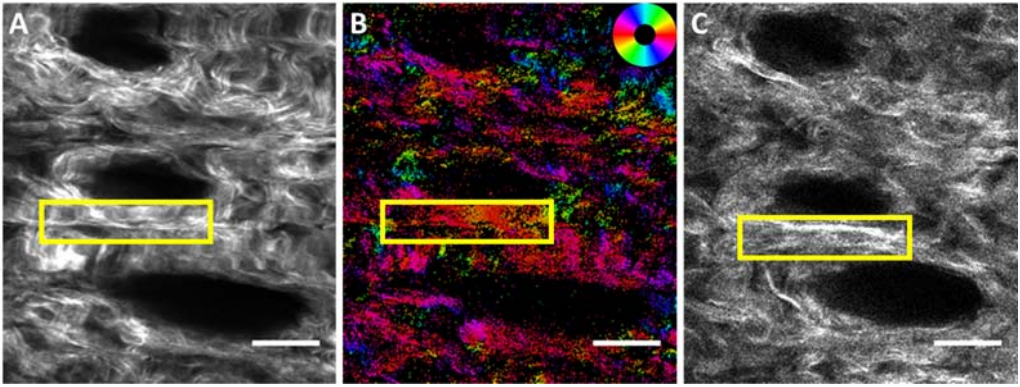

**Figure 2**. Zoomed-in images of a region with sample motion (*ex vivo* skin dermis of WT4 mouse at 1.05 stretch ratio). A: summation of all P-SHG images acquired at various linear incident polarizations; B: processed orientation image obtained from P-SHG acquisition; C: SHG image acquired by use of circular polarization. Note that the last image is slightly different from the first one because it was acquired afterwards and collagen reorganized. Scale bar: 20 µm.

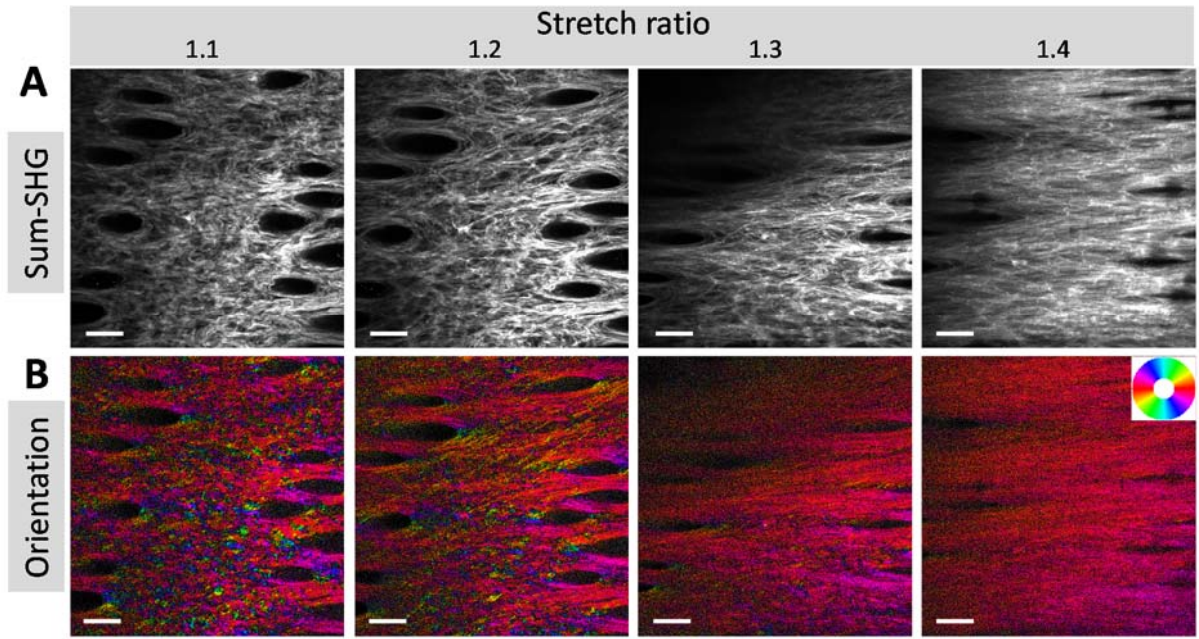

**Figure 3**. P-SHG imaging at increasing stretch ratio: 1.1, 1.2, 1.3 and 1.4 in skin sample WT15. A: Sum-SHG images obtained as the summation of all P-SHG images obtained for the different linear incident polarizations. B: Orientation maps processed from the P-SHG images and plotted using the HSV look-up table: H corresponds to the orientation shown in the inset, and V corresponds to $R^2$: V=1 for $R^2 \geq 0.5$ and V$\in$[0;1] for $R^2 \in$[0;0.5]. Scale bars: 50 μm.

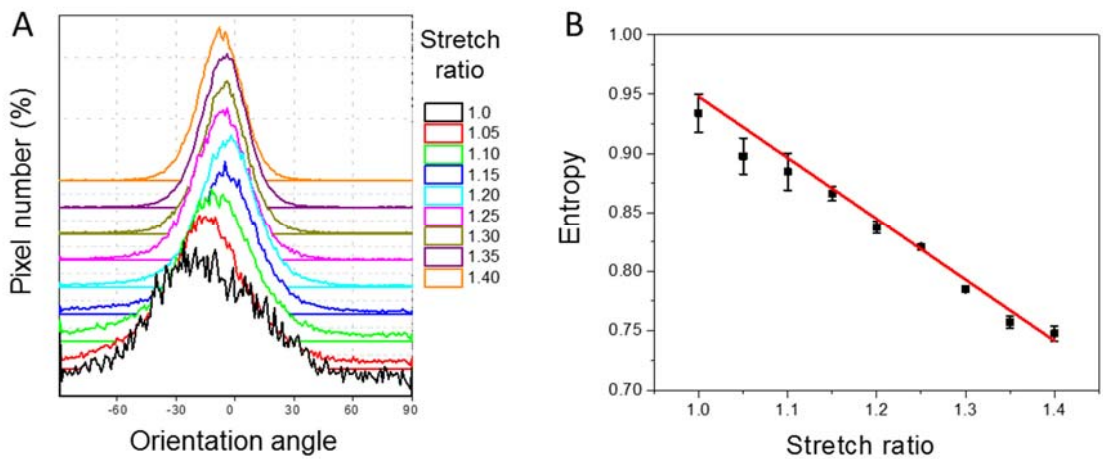

**Figure 4**. Reorganization of the collagen fibers network upon stretching in skin sample WT15 (same as in Figure 3). A: Orientation histograms obtained from P-SHG orientation maps using 1° bins at increasing stretch ratios. The baseline is shifted upwards for increasing stretch ratios. B: Variation of the normalized entropy of the orientation distributions in A as a function of the stretch ratio. The black squares correspond to the experimental data: mean ± SD calculated on the 6 2D P-SHG images exhibiting the maximal signal; the solid line corresponds to linear fitting.

**Table 1**. Measurements of dermal collagen reorganization upon stretching for 8 wild-type mice using morphological analysis and P-SHG method. The slope of the entropy decreases upon stretching, and the R² of the linear fit are provided for every mouse using both methods, either for 10° angular bins (morphological filtering and P-SHG) or for 1° angular bins (only P-SHG). The mean ± SEM of every parameter is also calculated.

| Mice | P-SHG - 1° bins | | P-SHG - 10° bins | | Morph. filtering | | Ratio | | |
|---|---|---|---|---|---|---|---|---|---|
| | slope | R² | slope | R² | slope | R² | P1°/M | P10°/M | P10°/P1° |
| WT4 | -0.31 | 0.96 | -0.55 | 0.96 | -0.05 | 0.84 | 6.33 | 11.19 | 1.77 |
| WT5 | -0.81 | 0.98 | -1.46 | 0.97 | -0.15 | 0.85 | 5.40 | 9.73 | 1.80 |
| WT6 | -0.50 | 0.96 | -0.88 | 0.96 | -0.18 | 0.96 | 2.76 | 4.89 | 1.77 |
| WT7 | -0.57 | 0.98 | -1.01 | 0.99 | -0.12 | 0.98 | 4.93 | 8.73 | 1.77 |
| WT9 | -0.36 | 0.83 | -0.61 | 0.85 | -0.10 | 0.70 | 3.58 | 6.06 | 1.69 |
| WT13 | -0.55 | 0.96 | -0.97 | 0.97 | -0.14 | 0.96 | 3.90 | 6.83 | 1.75 |
| WT14 | -0.73 | 0.96 | -1.28 | 0.96 | -0.17 | 0.95 | 4.41 | 7.73 | 1.75 |
| WT15 | -0.52 | 0.95 | -0.91 | 0.95 | -0.12 | 0.80 | 4.41 | 7.77 | 1.76 |
| **Mean** | -0.54 | 0.95 | -0.96 | 0.95 | -0.13 | 0.88 | 4.47 | 7.86 | 1.76 |
| SEM | 0.06 | 0.02 | 0.11 | 0.02 | 0.01 | 0.03 | 0.39 | 0.71 | 0.01 |

**Graphical Abstract**

An original fast polarization-resolved SHG microscope based on line-to-line switching of polarization using an electro-optical modulator is implemented to map the dynamic reorganization of the collagen network during mechanical assays. This system successfully highlights the collagen alignment along the traction direction in *ex vivo* murine skin dermis and enables quantitation of the linear increase of the collagen orientation distribution entropy as a function of the stretch ratio.

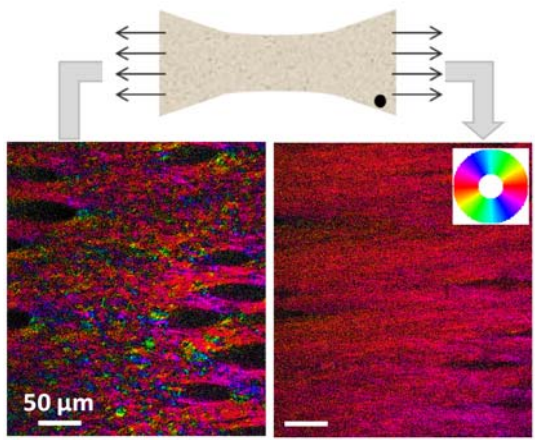